# Three-dimensional atomic positions and local chemical order of medium- and high-entropy alloys

Saman Moniri[1*], Yao Yang[1*], Yakun Yuan[1], Jihan Zhou[1], Long Yang[1], Fan Zhu[1], Yuxuan Liao[1], Yonggang Yao[2], Liangbing Hu[2], Peter Ercius[3], Jun Ding[4], and Jianwei Miao[1†]

[1]Department of Physics & Astronomy and California NanoSystems Institute, University of California, Los Angeles, CA 90095, USA. [2]Department of Materials Science and Engineering, University of Maryland, College Park, Maryland, 20742, USA. [3]National Center for Electron Microscopy, Molecular Foundry, Lawrence Berkeley National Laboratory, Berkeley, CA 94720, USA. [4]Center for Alloy Innovation and Design, State Key Laboratory for Mechanical Behavior of Materials, Xi'an Jiaotong University, Xi'an, China.

[*]These authors contributed equally to this work.

[†]Email: miao@physics.ucla.edu

**Medium- and high-entropy alloys (M/HEAs) mix multiple principal elements with near-equiatomic composition and represent a paradigm-shift strategy for designing new materials for metallurgy[1-8], catalysis[9-14], and other fields[15-18]. One of the core hypotheses of M/HEAs is lattice distortion[5,19,20], which has been investigated by numerical simulations, x-ray diffraction, and high-resolution transmission electron microscopy[21-26]. However, experimentally determining the three-dimensional (3D) local lattice distortion in M/HEAs remains a challenge. Additionally, the presumed random elemental mixing in M/HEAs has been questioned by atomistic simulations, energy dispersive x-ray spectroscopy (EDS), and electron diffraction, which suggest the existence of local chemical order in M/HEAs[27-33]. However, the 3D local chemical**

**order has eluded direct experimental observation since the EDS elemental maps integrate the composition of atomic columns along the zone axes[7,32,33], and the diffuse reflections/streaks in electron diffraction of M/HEAs may originate from planar defects instead of local chemical order[34]. Here, we determine the 3D atomic positions of M/HEA nanocrystals using atomic electron tomography[35], and quantitatively characterize the local lattice distortion, strain tensor, twin boundaries, dislocation cores, and chemical short-range order (CSRO) with unprecedented 3D detail. We find that the local lattice distortion and strain tensor in the HEAs are larger and more heterogeneous than in the MEAs. We observe CSRO-mediated twinning in the MEAs, that is, twinning occurs in energetically unfavoured CSRO regions but not in energetically favoured CSRO ones. This observation confirms the atomistic simulation results of the bulk CrCoNi MEA[27,29] and represents the first experimental evidence of correlating local chemical order with structural defects in any material system. We expect that this work will not only expand our fundamental understanding of this important class of materials, but also could provide the foundation for tailoring M/HEA properties through lattice distortion and local chemical order.**

Strength and ductility are two important material properties, but are mutually exclusive in most materials[36]. Recent experiments have demonstrated that several M/HEAs can overcome this strength-ductility tradeoff[3-8,37-39]. The high strength of M/HEAs stems from the different elemental components acting as solutes[40], local chemical order[29] and heterogeneous lattice strain[25,26], which raise the energy barrier of dislocation motion[34]. While dislocation glide in face-centred cubic (fcc) metals leads to high ductility with reduced strength[36], twinning in M/HEAs provides a distinct plasticity mechanism that obstructs dislocation motion and gains strength while remaining ductile[41]. This twinning-induced, simultaneous increase of strength and ductility in M/HEAs has spurred both mechanistically-driven and property-discovering investigations for structural applications[3,7,37-39,42]. However, the degree and influence of atomic segregation and CSRO on twin formation remain unclear in both M/HEAs and other alloys. Recent atomistic simulations have indicated a link between CSRO and twinning in MEAs[27,29], but there is no experimental evidence. Furthermore, CSRO, lattice distortion, and surface strain strongly affect the catalytic activity of M/HEAs[9-14]. Unlike conventional catalysts, M/HEAs confine different elements into the same lattice, which distorts the lattice structure and induces strain[12,13]. The lattice distortion and surface strain along with the chemical diversity of adsorption sites increase the activity, selectivity and durability of M/HEA catalysts[9,13,43-45] However, our current understanding of the lattice distortion, strain, and CSRO in M/HEA catalysts remains limited due to the dearth of 3D space and atomic scale information from diffraction, spectroscopy, electron microscopy, and atomistic simulations[9-14]. Here, we employ atomic electron tomography (AET) to determine the 3D atomic coordinates of NiPdPt-based M/HEA nanocrystals. We quantify the 3D lattice distortion, strain tensor, dislocations, twin boundaries and CSRO of the M/HEAs at the atomic scale. We also observe a direct link between CSRO and twinning in the MEAs. The observed CSRO in the nanocrystals agrees quantitatively with density functional theory (DFT)-calculated CSRO in the bulk, confirming that our conclusions are intrinsic and length scale-independent. Furthermore, the experimental 3D atomic coordinates are used as direct input to molecular dynamics (MD) simulations to calculate the twin formation energy of the MEAs, which agrees with our experimental observations of the twin positions.

**3D local lattice distortion and strain tensor in M/HEAs**

We choose NiPdPt-based M/HEA nanocrystals as a model system in this study for the following two reasons. First, they are catalytically active[14,46-48]. Second, they bypass the factors that have thus far hindered the direct and conclusive observation of CSRO in M/HEAs, namely the ambiguities in interpreting the diffuse reflections/streaks in the electron diffraction experiments[31,32,34] as well as the lack of scattering contrast in direct imaging. The latter is because AET has a power-law dependence on the atomic number[35]. The M/HEA nanocrystals were synthesized by heating metal precursors to ~2,000 K in ~50 ms, followed by rapid cooling at ~$10^5$ K/s (Methods)[49]. EDS maps confirm that the HEA nanocrystals consist of eight different elements (Extended Data Fig. 1). The AET experiments were performed with a scanning transmission electron microscope in the annular dark-field mode (Methods). We acquired tomographic tilt series from six MEA and four HEA nanocrystals (Extended Data Table 1). These nanocrystals were stable under the electron beam by examining the consistency of the images taken before, during and after the data acquisition. After image pre-processing, each tilt series was reconstructed by an advanced tomographic algorithm and the 3D atomic coordinates were traced, classified, and refined to produce an experimental atomic model (Methods). The experimental precision of the 3D atomic coordinates is 19.5 pm (Extended Data Fig. 2). While we resolve the atomic species in the MEA nanocrystals individually as Ni, Pd, and Pt, the eight elements in the HEA nanocrystals are classified into three types (Co and Ni as type 1; Ru, Rh, Pd, and Ag as type 2; Ir and Pt as type 3) because the atomic numbers of several elements are too close to be distinguished by AET[35]. The number of atoms and the atomic species / types of the ten M/HEAs are shown in Extended Data Table 1.

Figure 1a-d and Extended Data Fig. 3 show the experimental 3D atomic models of the ten M/HEA nanocrystals, exhibiting a single-phase fcc structure. To quantify the local lattice distortion of the M/HEAs, we compare each atom and its nearest neighbours with a reference fcc lattice to determine the 3D atomic displacement (Methods). Figure 1e-h shows the 3D atomic displacement for four representative MEAs and HEAs, termed MEA-1, MEA-2, HEA-1, and HEA-2, respectively. The mean and standard deviation of the atomic displacement of the four nanocrystals are 0.23±0.11 Å, 0.26±0.12 Å, 0.29±0.12 Å, and 0.37±0.12 Å, respectively (Fig. 1i-l), indicating that the HEAs have larger local lattice distortion than the MEAs. From the experimental 3D atomic coordinates, we also calculate the local strain tensor of the M/HEAs using a method described previously[50]. Figure 2 shows the six components of the local strain tensor for MEA-1, MEA-2, HEA-1 and HEA-2, in which the compressive, tensile, and shear strain range from -8% to +8%. Careful examination of Fig. 2 indicates the six strain tensor components of the HEAs are more heterogeneous than those of the MEAs, which is consistent with the observation of larger local lattice distortion in the HEAs (Fig. 1e-l).

Next, we quantitatively characterize twins and dislocations in the M/HEAs. Among the ten M/HEAs, three are twin-free, two have a single twin, four have double twins, and one has a grain boundary and double twins (Methods, Fig. 1a-d, Extended Data Fig. 3). The abundant presence of the single and double twins in M/HEAs is different from conventional nanocrystals, which have fewer twins except for decahedral multiply twinned nanocrystals[51,52]. Compared to the MEAs, the HEAs have more diffuse twin boundaries with each boundary spreading to the neighbouring atomic layers (Extended Data Fig. 4a-e). We also observe that the HEAs are more prone to have dislocations than the MEAs. Extended Data Fig. 4f-i shows the cores of several Shockley partial dislocations and screw dislocation in the M/HEAs and the corresponding Burgers vectors.

Determining the 3D atomic structure and measuring the lattice distortion, strain, twins and dislocations of M/HEAs not only answer some fundamental questions in materials science, but also provide key insights for the rational design of M/HEA-based catalysts[14]. We have recently demonstrated that using the experimental, non-relaxed atomic coordinates from AET as input to first-principles calculations can provide more accurate prediction of electronic properties than conventional calculations using relaxed structures[53]. M/HEA catalysis is an emerging field that has opened exciting opportunities for both theoretical and experimental investigations, particularly for multi-step reactions[14]. Thus, the conclusions from our 3D atomic structure analysis are applicable to both the structural and the functional applications of M/HEAs.

**Experimental observation of CSRO-mediated twinning**

To quantify the local chemical order in the M/HEAs, we compute the CSRO parameters ($\alpha_{ij}$) between each atom and its nearest neighbours[54,29] (Methods). For pairs of the same species / types ($i = j$), a positive $\alpha_{ii}$ means a tendency to segregate, and a negative $\alpha_{ii}$ does the opposite. For pairs of different species / types ($i \neq j$), a negative $\alpha_{ij}$ means in favour of inter-mixing, and a positive $\alpha_{ij}$ does the opposite. Fig. 3a, b and Extended Data Fig. 5a-d show the 3D distribution of the six CSRO parameters ($\alpha_{\text{NiNi}}$, $\alpha_{\text{PdPd}}$, $\alpha_{\text{PtPt}}$, $\alpha_{\text{NiPd}}$, $\alpha_{\text{NiPt}}$, and $\alpha_{\text{PdPt}}$) of the twin-free MEA-1 nanocrystal. The 3D distribution is heterogeneous with the formation of pockets of local chemical order, indicating that CSRO can propagate from the angstrom- to the nanometer-scale. To quantify the local chemical order, we average every CSRO parameter for each atomic layer along the [111] direction (Fig. 3c, d, Extended Data Fig. 5e-h). We observe that all the average $\alpha_{\text{NiPt}}$ values are negative and the majority of the average $\alpha_{\text{PdPt}}$ values are positive, indicating the tendency of the inter-mixing between Ni and Pt atoms and the separation between Pd and Pt atoms. To validate our experimental observations, we perform DFT-based calculations to predict energy-favoured CSRO in a twin-free bulk NiPdPt MEA (Methods). Figure 3i shows the histogram of the six average CSRO parameters of the twin-free MEA between the DFT calculations and experimental observations, confirming a favourable bonding between Ni and Pt atoms and an unfavourable bonding between Pd and Pt atoms in twin-free MEAs.

As a comparison, we calculate the six CSRO parameters for the double-twinned MEA-2. Figure 3e, f and Extended Data Fig. 5i-l show the 3D distribution of the six parameters, in which the yellow planes represent the twin boundaries. The CSRO of the double-twinned MEA are more heterogeneous than that of the twin-free MEA. Some pockets of local chemical order are connected to each other to form elongated structures that extend over a few nanometers. Figure 3g, h and Extended Data Fig. 5m-p show the histograms of the six average CSRO parameters of each atomic layer parallel to the twin boundaries. A majority of the $\alpha_{\text{NiPt}}$ and $\alpha_{\text{PdPt}}$ values are positive and negative, respectively (Fig. 3g, h), which are the opposite of the $\alpha_{\text{NiPt}}$ and $\alpha_{\text{PdPt}}$ of the twin-free MEA-1 (Fig. 3c, d). The average $\alpha_{\text{NiPt}}$ and $\alpha_{\text{PdPt}}$ values of the double-twinned MEA are the reverse of those of the twin-free MEA as well as the DFT calculations of a bulk twin-free MEA (Fig. 3i). These observations indicate that the separation of the favourable atomic species (Ni and Pt) and the inter-mixing of the unfavorable atomic species (Pd and Pt) facilitate the formation of the twins, which are further corroborated by the analysis of the other double-twinned MEA nanocrystal (Extended Data Figs. 6).

To investigate the impact of CSRO on the twin formation energy ($E_{TF}$) of the MEAs, we use the experimental 3D atomic coordinates and species as direct input to MD simulations and calculate $E_{TF}$ as a function of the twin position (Methods). We first apply this approach to a single-twinned MEA. Figure 4a-e shows the change of $E_{TF}$ by moving the twin from the $0^{th}$ to the $10^{th}$ atomic layer along the [111] direction, where the $0^{th}$ layer means twin-free. Four representative atomic configurations are shown in Fig. 4a-d with the twin marked in yellow. We observe that $E_{TF}$ changes from negative to positive when the twin is moved from atomic layer 5 to 6 (Fig. 4e). The experimentally determined twin position is in layer 5 (yellow bar), which is next to the minimum $E_{TF}$ in layer 4. Next, we use the experimental 3D atomic coordinates of the double-twinned MEA-2 as input to MD simulations to calculate $E_{TF}$ as a function of the twin separation. While fixing one twin, we move the other twin layer-by-layer along the [111] direction and compute the corresponding $E_{TF}$ (Fig. 4f-i). We find that $E_{TF}$ changes from negative to positive between a twin separation of 5 and 6 atomic layers. The experimentally determined twin separation is 5 layers (yellow bar in Fig. 4i), which is next to the minimum $E_{TF}$ with a twin separation of 4 layers. We analyze the other double-twinned MEA nanocrystal and obtain consistent results (Extended Data Fig. 7). All these observations show the correlation between the CSRO and twinning in the MEAs, that is, energetically-unfavourable CSRO lowers the $E_{TF}$ and mediates the formation of the twins.

Compared to the MEAs, the HEAs exhibit greater local chemical fluctuations. Extended Data Figs. 8a-f and 9a-f show the 3D distribution of the six CSRO parameters for a twin-free and a double-twinned HEA, which are more heterogeneous than those of the MEAs (Fig. 3a, b, e, f, Extended Data Fig. 5a-d, i-l, 6a-f). The increase of the chemical complexity from the three-element MEAs to the eight-element HEAs leads to a larger distortion of the twin geometry as manifested in the formation of atomic steps along the twins of both the MEAs and HEAs (Extended Data Fig. 4b-e), with the latter containing multiple such steps that disrupt stacking of the neighbouring atomic planes. To examine the correlation between CSRO and twinning in the HEAs, we calculate the six average CSRO parameters of the atomic layer along the [111] direction for a twin-free HEA (Extended Data Fig. 8g-l). In comparison, the six average CSRO parameters for a double-twinned HEA are shown in Extended Data Fig. 9g-l. We observe the reverse of $\alpha_{13}$ between the twin-free and double-twinned HEAs, indicating that the separation of the favourable atomic types 1 and 3 facilitates the formation of the twins in the HEAs (Extended Data Fig. 9m). Because our current AET experiment can only classify the eight elements in the HEAs into three types, we cannot calculate the $E_{TF}$ from the experimental 3D atomic coordinates and types of the HEAs as in the case of the MEAs.

**Conclusions and outlook**

Lattice distortion and CSRO are two fundamental features that strongly affect the unique properties of M/HEAs. Despite significant insights from various experiments[21-32], direct 3D structure information remains elusive. Here we overcome this major limitation by advancing AET to determine the 3D atomic positions of M/HEAs and quantitatively characterize their local lattice distortion, strain tensor, dislocation cores, and CSRO in three dimensions. Compared to the MEAs, the HEAs exhibit greater local structural and chemical heterogeneity, resulting in more complex local strain, diffuse twin boundaries and dislocations. We also observe CSRO-mediated twinning in the MEAs, which corroborates the DFT calculations of the bulk NiPdPt MEA (Fig. 3i, Methods) and the atomistic simulations of the bulk CrCoNi MEA[27,29]. This demonstrates that CSRO-mediated twinning is a general mechanism and independent of the length scale. Owing to the tunability of CSRO during materials manufacturing, our 3D atomic-scale insights into the correlation between CSRO and twinning could expand the horizon for the design of M/HEAs and other alloys with targeted structure-property relationships. In catalysis, M/HEAs have shown performance improvements over conventional alloys for various multi-step reactions[12,14], including ammonia oxidation/decomposition[9,49], carbon dioxide reduction[43,44], and methane combustion[55]. Additionally, M/HEA catalysts possess near-continuum adsorbate binding energies and greater structural stability[12,14,48]. Thus, determining the 3D atomic structure of M/HEA catalysts and measuring their 3D local lattice distortion and strain tensor open the door for their rational design in a largely untapped range of compositions and structures. Looking forwarding, AET can be combined with EDS to map out the complete 3D elemental distribution of any M/HEA. We expect that the ability to determine the 3D atomic structure and local chemical order of M/HEAs will extend to both the metallurgical community's pursuit of superior strength-ductility combinations, and the catalysis community's quest toward optimized surface adsorption energies.

**Figures and Figure legends**

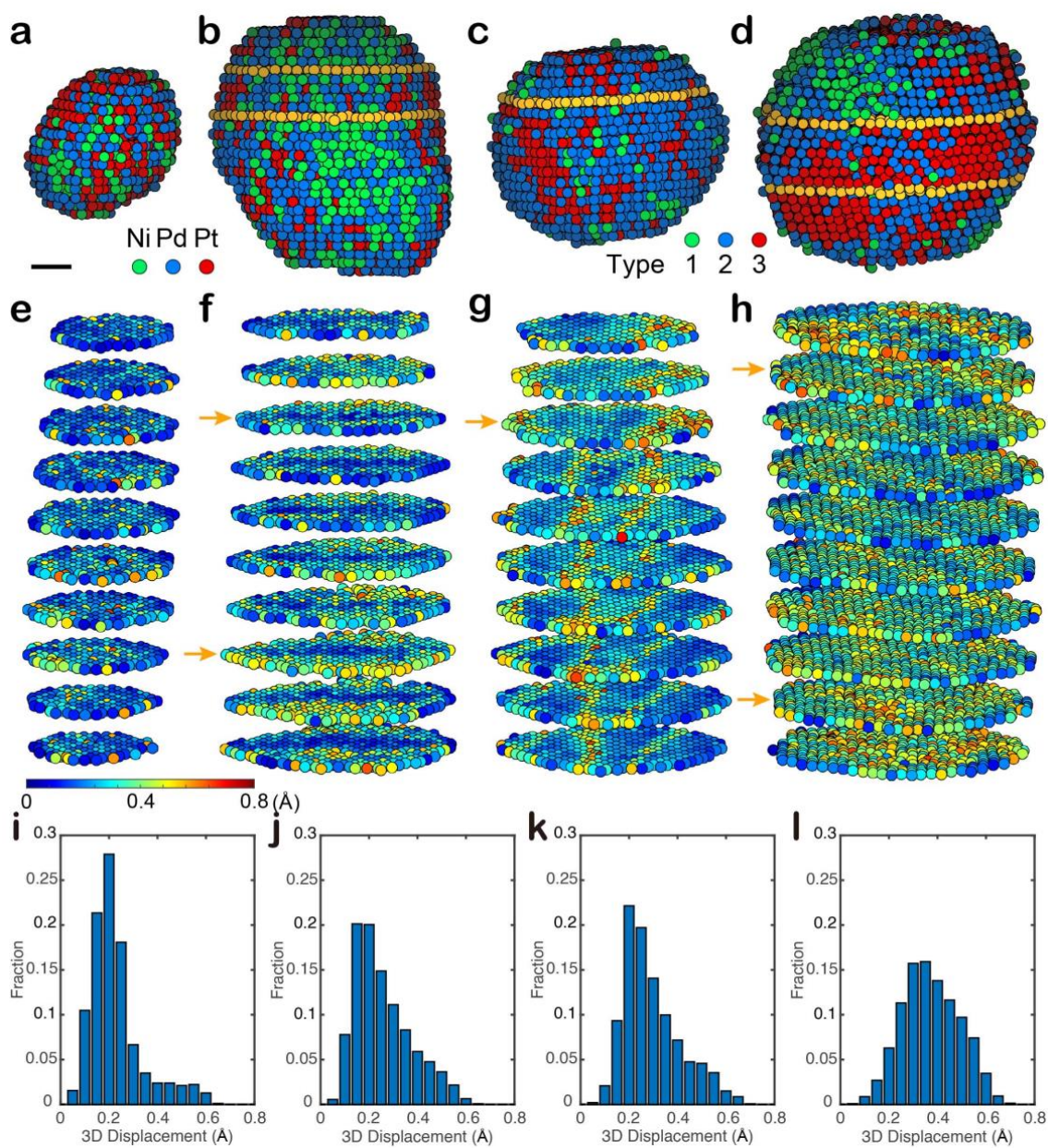

**Fig. 1 | The 3D atomic structure and lattice distortion of M/HEA nanocrystals. a-d**, Experimental atomic models of two MEAs and two HEAs, named MEA-1 (**a**), MEA-2 (**b**), HEA-1 (**c**), and HEA-2 (**d**), respectively, in which the yellow circles represent the atoms along the twin boundaries. **e-h**, Atomic layer-by-layer visualization of the 3D

displacement of MEA-1, MEA-2, HEA-1, and HEA-2, respectively, where the yellow arrows point to the twin boundaries. **i-l**, The distributions of the 3D atomic displacement of MEA-1, MEA-2, HEA-1, and HEA-2, respectively. Scale bar, 1 nm.

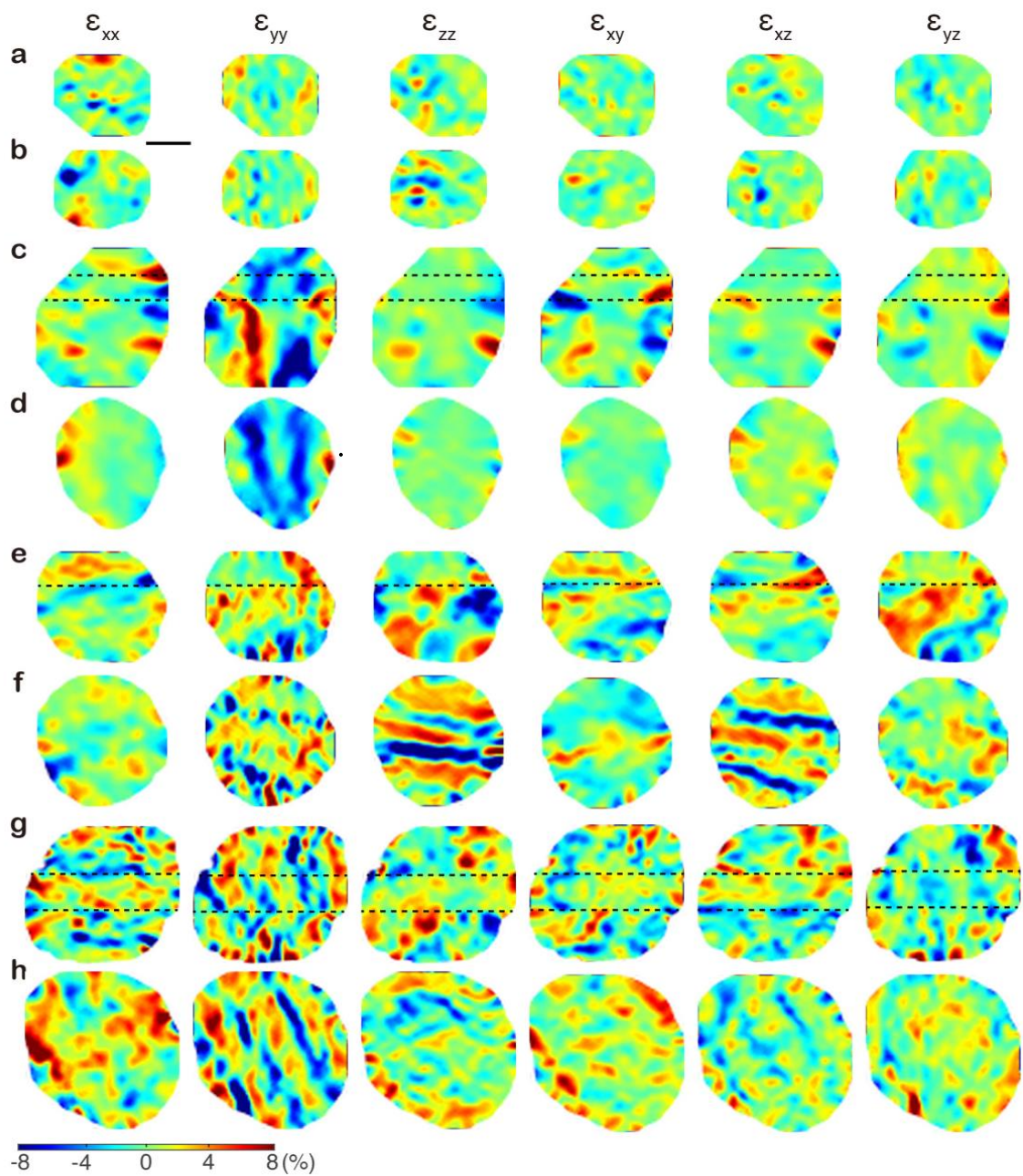

**Fig. 2 | 3D strain tensor measurements of the M/HEA nanocrystals. a-h**, Six components of the local strain tensor of the two perpendicular atomic layers in MEA-1 (**a**, **b**), MEA-2 (**c**, **d**), HEA-1 (**e**, **f**), and HEA-2 (**g**, **h**), in which the dashed lines represent

the twin boundaries. (**a**, **c**, **e**, and **g**) are along the $[1\bar{1}0]$ direction and (**b**, **d**, **f**, and **h**) along the [111] direction. Scale bar, 2 nm.

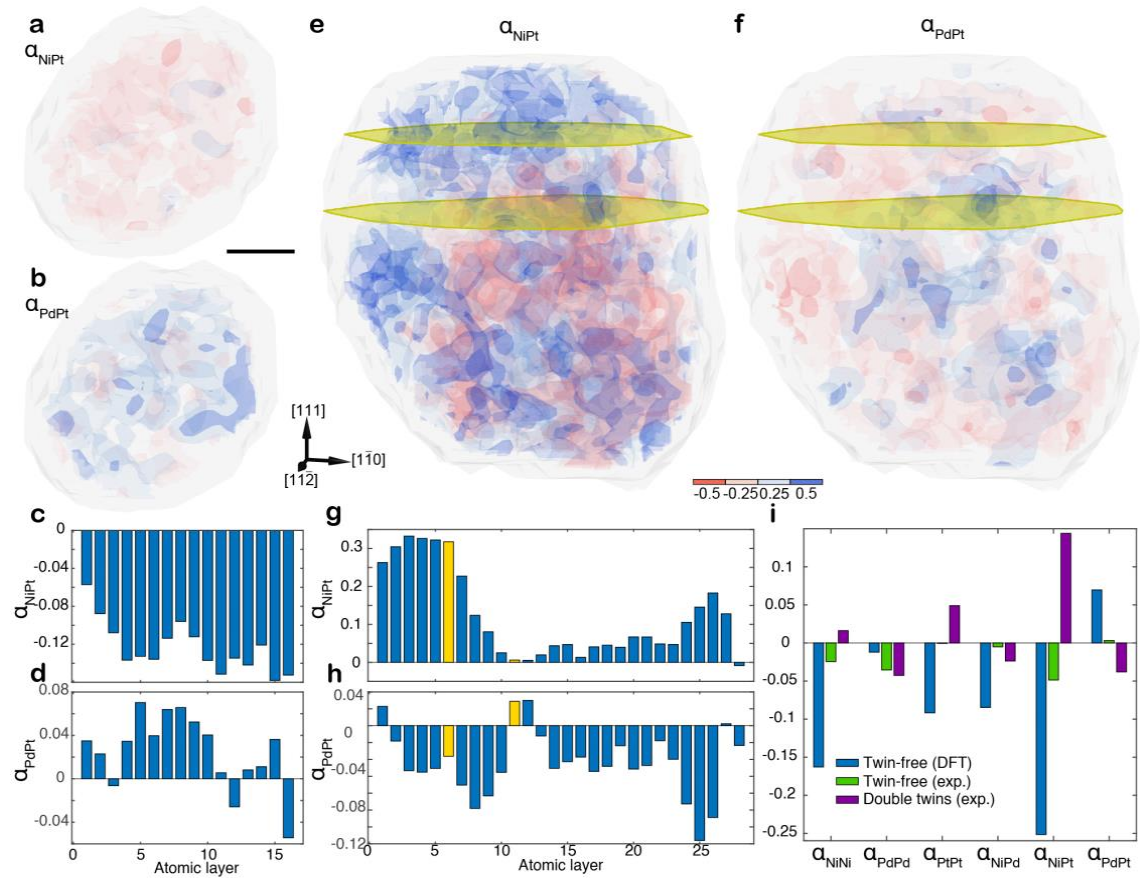

**Fig. 3 | Experimental observation of the correlation between CSRO and twinning in MEAs. a**, **b**, 3D distribution of $\alpha_{\text{NiPt}}$ and $\alpha_{\text{PdPt}}$ in the twin-free MEA-1. **c**, **d**, Histogram of the average $\alpha_{\text{NiPt}}$ and $\alpha_{\text{PdPt}}$ for each atomic layer along the [111] direction, indicating the tendency of the inter-mixing between Ni and Pt atoms and the separation between Pd and Pt atoms. **e**, **f**, 3D distribution of $\alpha_{\text{NiPt}}$ and $\alpha_{\text{PdPt}}$ in the double-twinned MEA-2 (the twins marked with yellow planes), exhibiting more heterogeneous CSRO than the twin-free MEA-1 (**a** and **b**). **g**, **h**, Histogram of the average $\alpha_{\text{NiPt}}$ and $\alpha_{\text{PdPt}}$ values for each atomic layer along the [111] direction (the yellow bars indicate the twin positions), which are the opposite of the $\alpha_{\text{NiPt}}$ and $\alpha_{\text{PdPt}}$ for the twin-free MEA-1 (**c**, **d**). **i**, Histogram of

the six average CSRO parameters of a DFT-calculated bulk MEA, the twin-free MEA-1, and the double-twinned MEA-2.

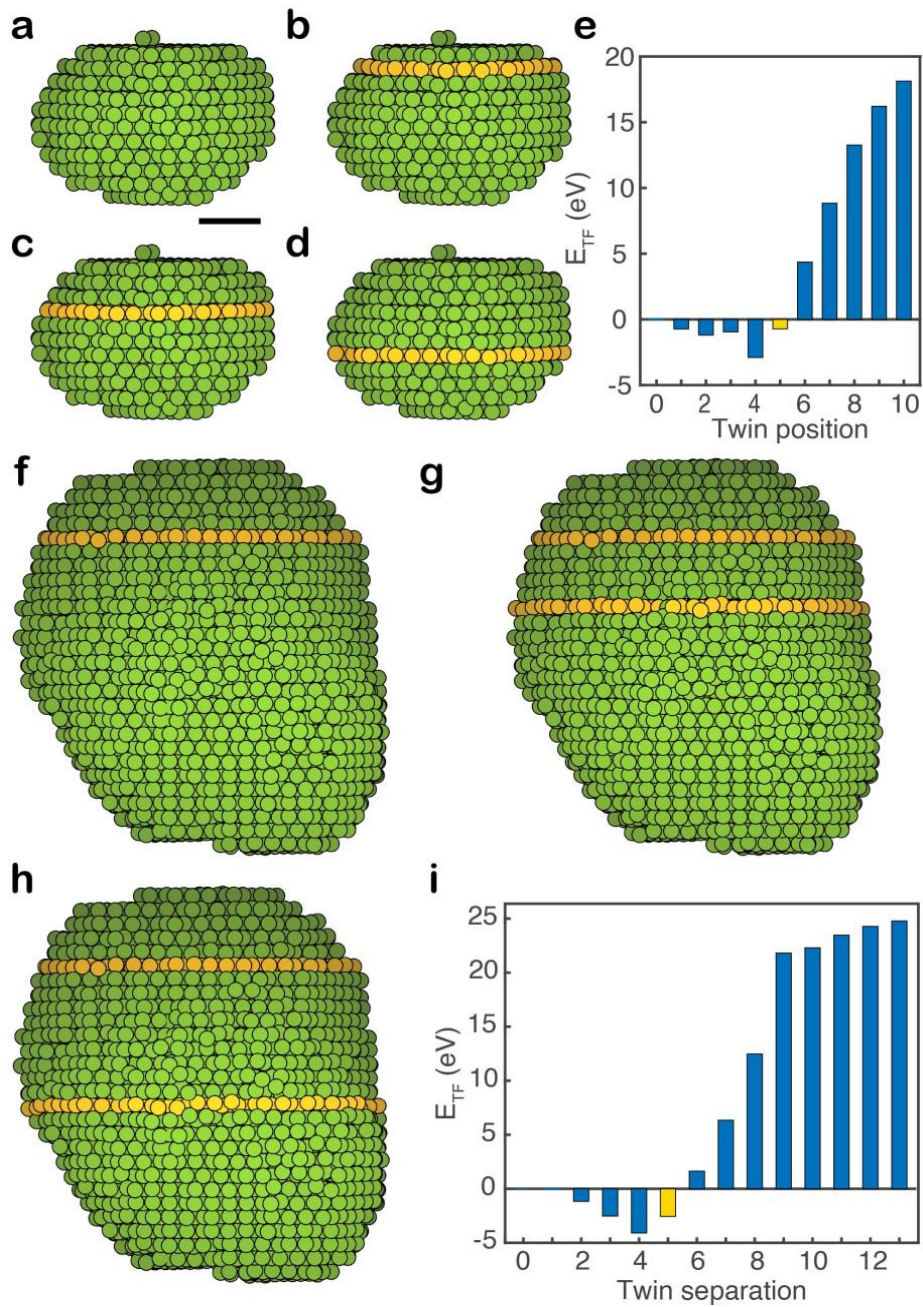

**Fig. 4 | The twin formation energy ($E_{TF}$) calculated from the experimental 3D atomic coordinates and species of the MEAs**. **a-d**, Calculation of $E_{TF}$ of a single-twinned MEA by moving the twin (yellow circles) from the $0^{th}$ (top) to the $10^{th}$ atomic layer along the [111] direction, with four representative atomic configurations showing the twin in layer

0 (i.e. twin-free) (**a**), 3 (**b**), 5 (**c**) and 8 (**d**). **e**, Histogram of $E_{TF}$ as a function of the twin position, in which $E_{TF}$ changes from negative to positive from atomic layer 5 to 6. The experimentally determined twin is in layer 5 (yellow bar), which is next to the minimum $E_{TF}$ in layer 4. **f-h**, Calculation of $E_{TF}$ of the double-twinned MEA by fixing one twin (yellow circles in **f**) and moving the other twin along the [111] direction, where three representative atomic configurations show a twin separation by 0 (i.e., a single twin) (**f**), 5 (**g**), and 10 atomic layers (**h**). **i**, Histogram of $E_{TF}$ as a function of the twin separation, in which $E_{TF}$ changes from negative to positive between a twin separation of 5 and 6 layers. The experimentally determined twin separation is 5 atomic layers (yellow bar), which is next to the minimum $E_{TF}$ with a twin separation of 4 layers.

## METHODS

**Specimen synthesis and preparation.** The M/HEA samples were prepared via conventional carbothermal shock method[49]. Metal precursors in ethanol (0.05 M) were first mixed into the desired multielement composition, i.e., NiPdPt and CoNiRuRhPdAgIrPt. The mixed precursors were then dipped onto carbon heater for Joule heating, where reduced graphene oxide was used as the substrate for dispersion and stabilization. After room temperature drying, metal precursors on the reduced graphene oxide were rapidly heated to a high temperature (~2,000 K for ~50 ms) by Joule heating followed by rapid cooling (~$10^5$ K/s), which produced the M/HEA nanocrystals.

**AET data acquisition**

The AET experiments of ten M/HEA nanocrystals were performed using the TEAM 0.5 microscope with the TEAM stage at the National Center for Electron Microscopy, operated in ADF-STEM mode at 200 or 300 kV (Extended Data Table 1). At each tilt angle, a nearby nanocrystal or a sample feature was used as a fiducial to align and focus the image, thereby reducing the unnecessary exposure of the region of interest to the electron beam[56]. To minimize the drift distortion and electron dose at each tilt angle, three to four sequential images were taken with a dwell time of 3 μs. The total electron dose of each tilt series was optimized to be between $7.4\times10^5$ and $8.5\times10^5$ e$^-$/Å$^2$ to reduce the beam damage (Extended Data Table 1).

For each sample, we confirmed the structural stability upon beam exposure by comparing images before, during, and after the acquisition of the tilt series.

**Image pre-processing**

A multi-pronged image pre-processing protocol was performed on each AET dataset as outlined below.

i) Drift correction. To compensate for sample drift during data acquisition, we collected three to four images at each tilt angle, computed the cross-correlation coefficient between the images and identified the relative drift vectors by the maximum cross-correlation. We used a step size of 0.1 pixel as the drift in typical ADF-STEM images is smaller than one pixel. We applied the drift correction to each image along the slow scan direction, and corrected for it by interpolating the raw images with drift-corrected pixel positions. The drift-corrected images were then averaged to form a single image per tilt angle.

ii) Image denoising. To remove the Poisson and Gaussian noise from the drift-corrected images, we used the block-matching and 3D filtering (BM3D) algorithm[57], which has been successful in denoising experimental AET datasets[58,59]. To optimize the BM3D denoising parameters for each dataset, we first estimated the level of each noise type (Poisson and Gaussian) in the image stack. Then, these noise levels were added to several simulated ADF-STEM images of model nanocrystals with similar size and elemental distribution as the experimental data. The denoising parameters leading to the maximum cross-correlation between the simulated images and the experimental images were applied to the experimental data.

iii) Background subtraction. For each denoised image, we used Otsu thresholding in MATLAB to generate a mask of the nanocrystal, slightly larger than its boundary. Using the background outside the masked region, we performed Laplacian interpolation to estimate the background level inside the masked zone and subtracted it from the denoised image[58].

iv) Image alignment. The alignment of the background subtracted images was achieved by the common line method along the tilt axis and the centre of mass method perpendicular to the tilt axis, which have been demonstrated to align experimental tilt series with sub-pixel accuracy[56,60].

**Tomographic reconstruction**

Each pre-processed dataset was reconstructed by the REal Space Iterative REconstruction (RESIRE) algorithm[61]. RESIRE iteratively minimizes the difference between the experimental and calculated projections of the sample using gradient descent. Through the incorporation of angular refinement and spatial alignment, RESIRE is superior to other tomographic reconstruction algorithms[61]. A typical RESIRE reconstruction converges after about 200 iterations. Following an initial reconstruction and an iterative process of angular refinement and spatial alignment, the background of the new set of images was re-evaluated and re-subtracted. The resulting images were then used to generate the final 3D reconstruction with RESIRE, after undergoing further angular refinement and spatial realignment.

**Determining the 3D atomic coordinates and species / types**

From the final 3D reconstructions, we determined the 3D atomic coordinates and species / types using the following procedure.

i) Each reconstruction underwent spline interpolation to produce a finer grid using an oversampling ratio of 3, from which all the local maxima were identified. Using a polynomial fitting method[58], we identified the positions of all the peaks (i.e., potential atoms) from a 0.8 Å × 0.8 Å × 0.8 Å volume around each peak. An initial list of the potential atoms was obtained by searching through all the fit peak positions with the constraint that the minimum distance between neighbouring peaks is 2 Å, given that all the interatomic distances in our samples are larger than this value.

ii) To remove the non-atoms from this list, we performed K-means clustering of the integrated intensity of the local volume around each potential atom position[58].

iii) By overlaying the 3D atomic positions in the updated list on the 3D reconstructions, we manually checked the atomic positions and corrected the fitting failure of a small fraction (< 1%) of unidentified or misidentified atoms. We note that such manual correction is routine for atom tracing and refinement in macromolecular crystallography[62].

iv) K-means clustering was used iteratively to classify the Ni, Pd and Pt atoms in MEAs and the three types of atoms in HEAs (Co and Ni as type 1; Ru, Rh, Pd and Ag as type 2; and Ir and Pt as type 3) based on the integrated intensity of the 0.8 Å × 0.8 Å × 0.8 Å volume around each atom position.

v) We performed local reclassification of all the atomic species / types. Each atom was defined to be at the centre of a 10 Å-radius sphere. The average intensity distribution of the three atomic species / types was computed within this sphere. We then computed the $L_2$ norm of the intensity distribution between the central atom and the average Ni, Pd and Pt or type-1, -2, and -3 atoms. The atom was assigned to the species / type with the smallest $L_2$ norm. After repeating this step for all the atoms, an initial experimental 3D atomic model of the sample was obtained.

vi) The 3D atomic coordinates of the initial model were refined by minimizing the error between the experimental and computed images using gradient descent as described elsewhere[63]. Convergence of the iterative process was verified by monitoring the $L_2$ norm error.

**Local lattice distortion**

The local lattice distortion is defined as the deviation of the experimental atomic positions in the M/HEA nanocrystals from those in a perfect fcc lattice. For each atom, its local lattice distortion ($\Delta d_i$) was calculated by,

$$\Delta d_i = \frac{1}{N}\sum_j |\mathbf{r}_{ji} - \mathbf{r}_{ji}^0| \qquad (1)$$

where $N$ is the number of the nearest neighbours of atom $i$, $\mathbf{r}_{ji}$ is the experimental 3D coordinates of the $j^{\text{th}}$ nearest-neighbour atom, and $\mathbf{r}_{ji}^0$ is the perfect fcc lattice structure that was aligned to $\mathbf{r}_{ji}$ based on the three Euler angles determined by a breadth-first search algorithm[58]. A cut-off of a quarter of the nearest-neighbour bond length was applied to eliminate the contribution from some surface atoms with a large deviation.

**CSRO parameters**

After identifying the nearest neighbours of each atom, we computed the CSRO parameters ($\alpha_{ij}$) between the central atom ($i$) and its nearest neighbours ($j$) by[54,29]

$$\alpha_{ij} = \frac{p_{ij} - c_j}{\delta_{ij} - c_j} \qquad (2)$$

where $p_{ij}$ denotes the average probability that a $j$-type atom is the nearest neighbour to an $i$-type atom, $c_j$ signifies the average concentration of $j$-type atoms, and $\delta_{ij}$ is the Kronecker delta function. To eliminate the boundary effect, we removed the surface atoms of each M/HEA nanocrystal from the calculation of the CSRO parameters. Following this procedure, we computed the six CSRO parameters for every atom in the M/HEA nanocrystal, which are bounded between -1 and +1. To obtain a local CSRO distribution such as those in Fig. 3, Extended Data Figs. 5, 6, 8 and 9, we interpolated the CSRO parameters onto 3D grids and convolved them with a Gaussian kernel. The width of the Gaussian kernel was determined by the first valley of the pair distribution function of the nanocrystal, corresponding to the first nearest-neighbour shell distance.

**Twin order parameter**

From the experimental 3D atomic coordinates, we fit the nearest neighbour atoms around each atom to a perfect fcc and hexagonal close-packed (hcp) lattice by the breadth-first search algorithm[58]. If a nearest neighbour atom has a deviation larger than a cut-off, we set its deviation to be the cut-off value. This step was to eliminate the effect of some surface atoms with a large deviation. The twin order parameter was calculated by[64]

$$\eta = \frac{d_{\text{fcc}} - d_{\text{hcp}}}{d_{\text{max}}} \qquad (3)$$

where $d_{\text{fcc}}$ and $d_{\text{hcp}}$ are the sum of the deviation of the nearest-neighbour atoms from a perfect fcc and hcp lattice, respectively, and $d_{\text{max}}$ is the maximum deviation. In this study, we chose a cut-off value of 0.75 Å. We also calculated $\eta$ using different cut-off values and obtained consistent results. We used $\eta$ to determine the twin boundaries in the M/HEA nanocrystals (Extended Data Fig. 4a-e), where $\eta$ = 1 and -1 represent a hcp (i.e., twinning) and fcc structure, respectively.

**DFT calculations**

DFT-based lattice Monte Carlo (MC) approach[27] was utilized to reveal the characteristic CSRO for the bulk NiPdPt MEA and bulk NiCoRuRhPdAgIrPt HEA. The 256-atom configurations were generated as a special quasi-random structure[65]. The MC simulations ran for 3,000 steps, corresponding to about 12 swap trials per atom, at an MC temperature of 600 K. Energy calculations were performed using the Vienna *ab initio* simulation package[66,67]. A plane wave cut-off energy was chosen at 380 eV, and the Brillouin zone integrations were performed using Monkhorst–Pack meshes[67] with a single k-point (Γ). It adopted projector augmented wave potentials[68] with the Perdew–Burke–Ernzerhof generalized-gradient approximation[69] for the exchange-correlation functional.

**Twin formation energy of the MEAs**

We calculated the twin-formation energy of the NiPdPt MEAs from the experimentally measured 3D atomic coordinates using the LAMMPS software package[70] with the empirical EAM potential[71]. The experimental 3D atomic coordinates of the MEA nanocrystals were extracted from the AET experiments. To eliminate the boundary effect, the surface atoms of each MEA nanocrystal were not used in the MD calculations. For the supercell, the shrink-wrapped non-periodic boundary conditions were imposed along all three directions. As illustrated in Fig. 4 and Extended Data Fig. 7, the migration of the twin boundary was carried out by gradually shifting the close-packed (111) atomic planes along the ⟨112⟩ direction by the Burgers vector of the Shockley partial $b_s = \frac{a}{6}\langle 11\bar{2}\rangle$. For the twin boundary at different position, their formation energy was calculated by the energy difference between the twinned and twin-free configurations after energy minimization.

**Data availability**

All the raw and processed experimental data will be immediately posted on an open-access repository (https://github.com) after the paper is published online.

**Code availability**

All the MATLAB source codes for the 3D reconstruction, atom tracing, and data analysis of this work will be immediately posted on an open-access repository (https://github.com) after the paper is published online.

**Acknowledgements** This work was primarily supported by the U.S. Department of Energy, Office of Science, Basic Energy Sciences, Division of Materials Sciences and Engineering under Award No. DE-SC0010378. The ADF-STEM imaging with TEAM 0.5 was performed at the Molecular Foundry, which is supported by the Office of Science, Office of Basic Energy Sciences of the U.S. DOE under Contract No. DE-AC02—05CH11231.

**Extended Data Figures and Table**

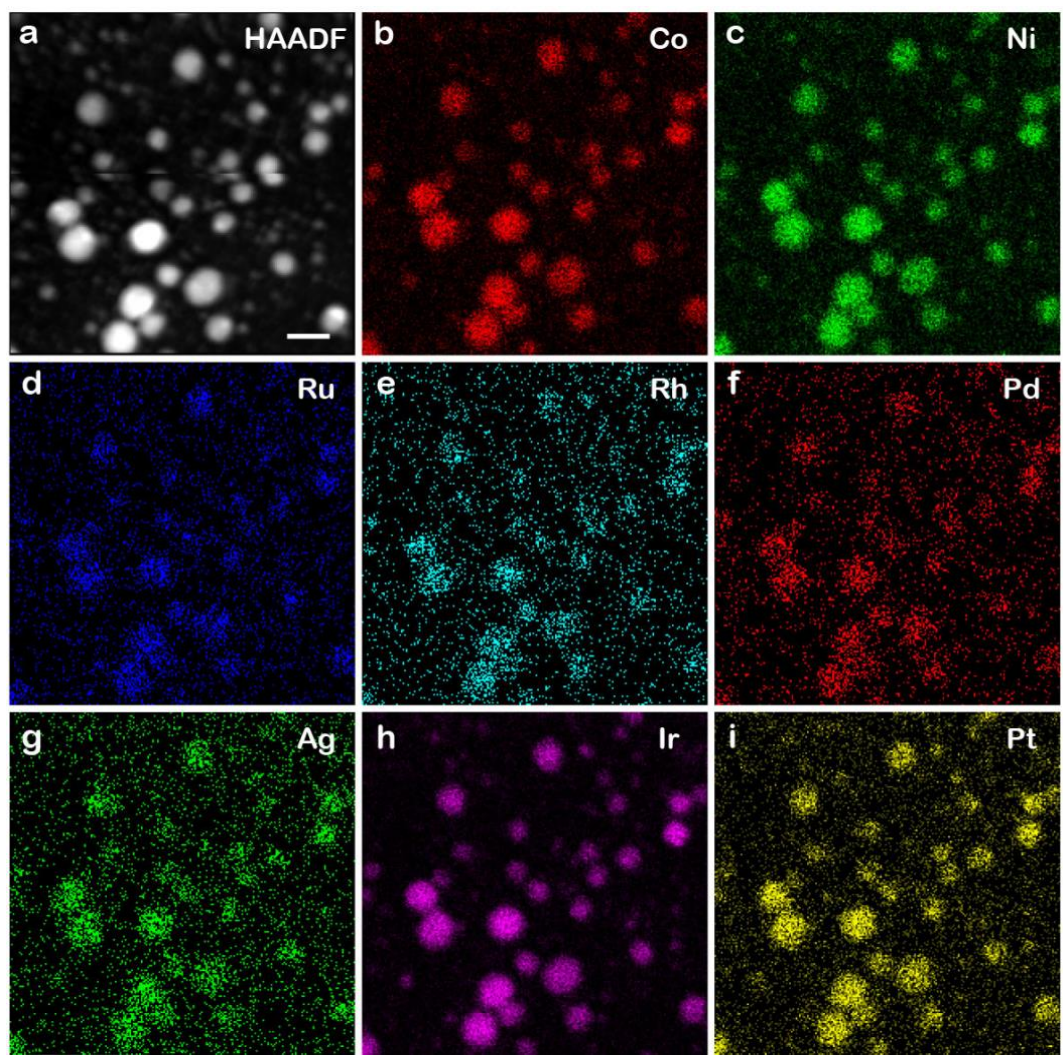

**Extended Data Fig. 1 | EDS maps of the HEA nanocrystals. a**, Low-resolution ADF-STEM image of the nanocrystals. **b-i**, EDS maps showing the distribution of Co (b), Ni (c), Ru (d), Rh (e), Pd (f), Ag (g), Ir (h) and Pt (i) in the nanocrystals. Scale bars, 20 nm.

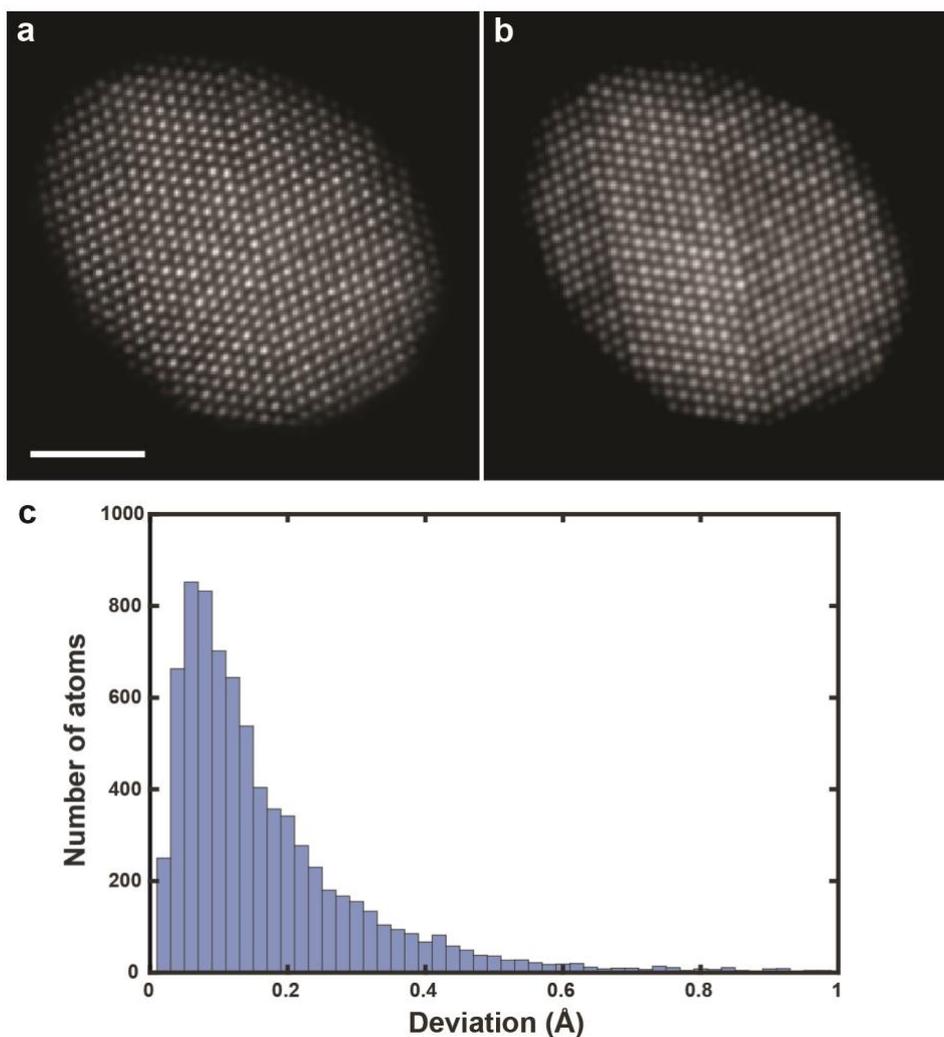

**Extended Data Fig. 2 | 3D precision estimation**. **a**, **b**, Comparison between a representative experimental (after denoising) (**a**) and a multi-slice calculated image (**b**) of MEA-3. The multi-slice images were convolved with a Gaussian function to match the incoherence effects in the experimental images. **c**, Histogram of the deviation of the 3D atomic positions between the experimental atomic model and that obtained from 55 multi-slice images. The root-mean-square deviation of the histogram is 19.5 pm. Scale bar, 2 nm.

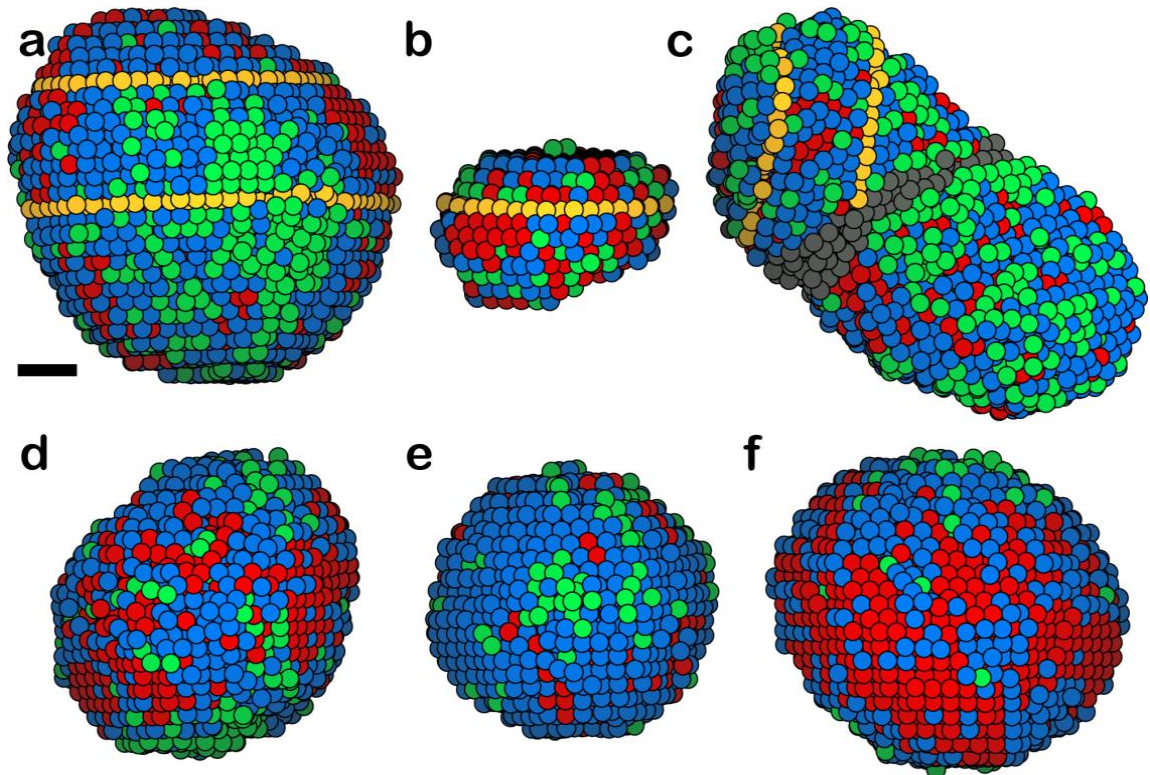

**Extended Data Fig. 3 | Experimental 3D atomic models of the other six M/HEA nanocrystals**. **a-f**, Experimental atomic models of four MEAs and two HEAs, named MEA-3 (**a**), MEA-4 (**b**), MEA-5 (**c**), MEA-6 (**d**), HEA-3 (**e**) and HEA-4 (**f**), in which the yellow circles represent the twin boundaries in (**a-c**) and grey circles indicate a grain boundary in (**c**). Scale bar, 1 nm.

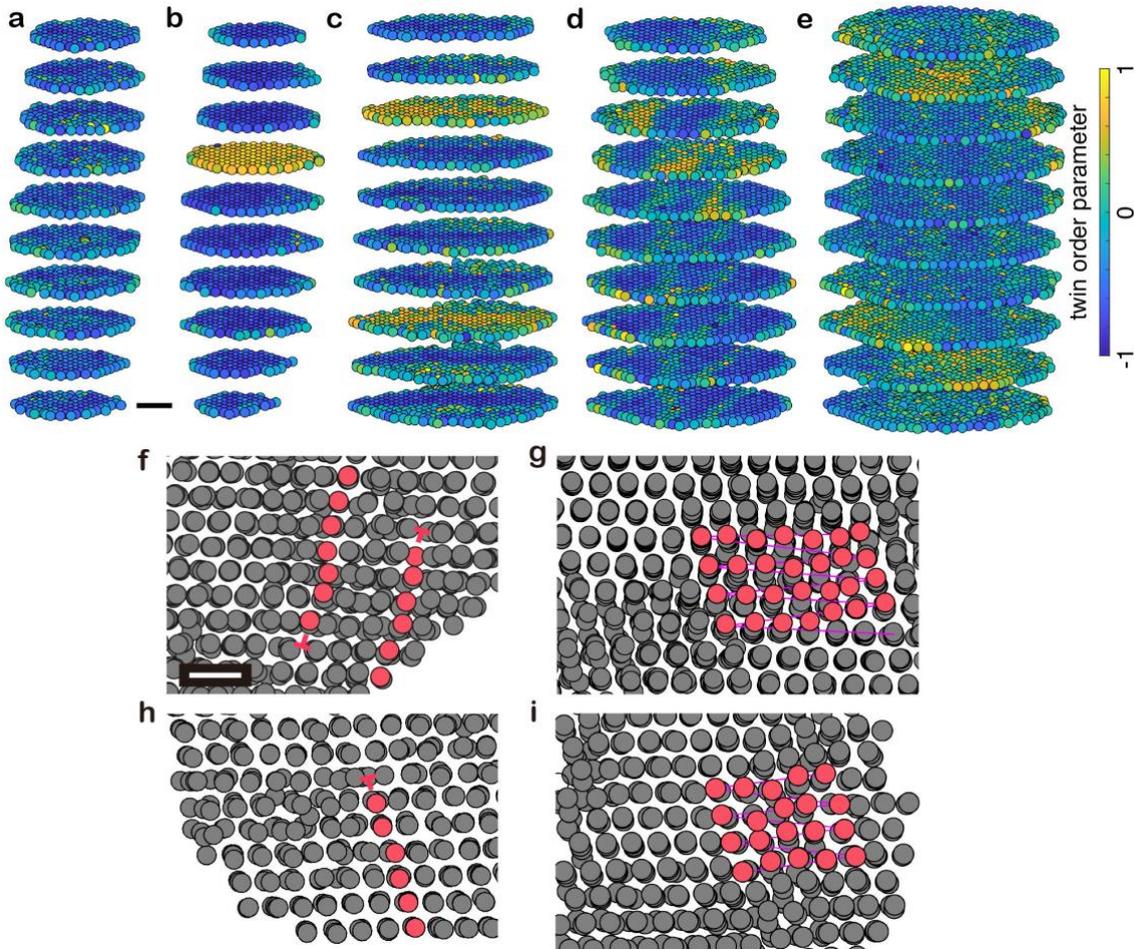

**Extended Data Fig. 4 | Twin boundaries and dislocations in the M/HEAs. a-e,** The twin boundaries of three representative MEAs (**a**, twin-free; **b**, single twin; **c**, double twins) and two HEAs (**d**, single twin; **e**, double twins), showing more diffuse twin boundaries in the HEAs with each boundary spreading to the neighbouring atomic layers. The twin order parameter of 1 and -1 represents a hcp (i.e., twinning) and fcc structure, respectively. **f,** Two Shockley partial dislocations in MEA-5 with opposite Burgers vectors $\frac{a}{6}[121]$ and $\frac{a}{6}[\bar{1}\bar{2}\bar{1}]$, as the gliding process was frozen near the boundary during the rapid cooling process of the nanocrystal. **g,** A screw dislocation in MEA-2 with the Burgers vector $\frac{a}{2}[110]$. **h,** A Shockley partial dislocation in HEA-4 with the Burgers vector $\frac{a}{6}[121]$, which exists near the boundary of the nanocrystal. **i,** A screw dislocation in HEA-3 with the Burgers vector $\frac{a}{2}[110]$. Scale bars, 1 nm (**a**); and 5 Å (**f**).

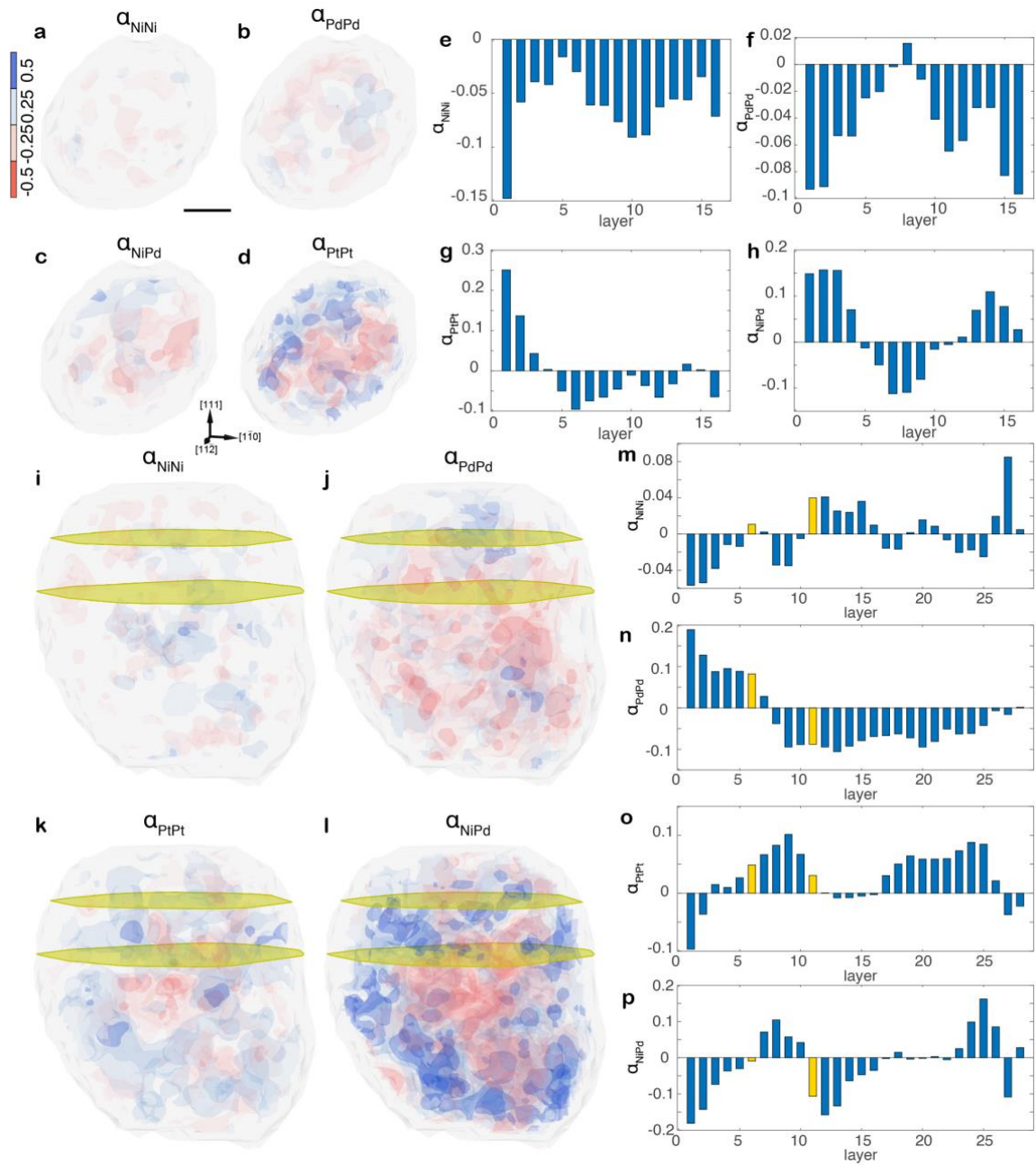

**Extended Data Fig. 5 | 3D distribution of the other four CSRO parameters of the twin-free MEA-1 and double-twinned MEA-2**. **a-d**, 3D distribution of $\alpha_{NiNi}$, $\alpha_{PdPd}$, $\alpha_{PtPt}$, and $\alpha_{NiPd}$ in MEA-1, showing the formation of local chemical-order pockets. **e-h**, Histogram of the average $\alpha_{NiNi}$, $\alpha_{PdPd}$, $\alpha_{PtPt}$, and $\alpha_{NiPd}$ for each atomic layer along the [111] direction in MEA-1. **i-l**, 3D distribution of $\alpha_{NiNi}$, $\alpha_{PdPd}$, $\alpha_{PtPt}$, and $\alpha_{NiPd}$ in MEA-2 (the twins labelled as yellow planes), exhibiting more heterogeneous CSRO than the twin-free MEA-1 (**a-d**). **m-p**, Histogram of the average $\alpha_{NiNi}$, $\alpha_{PdPd}$, $\alpha_{PtPt}$, and $\alpha_{NiPd}$ for each atomic layer along the [111] direction in MEA-2. Scale bar, 1 nm.

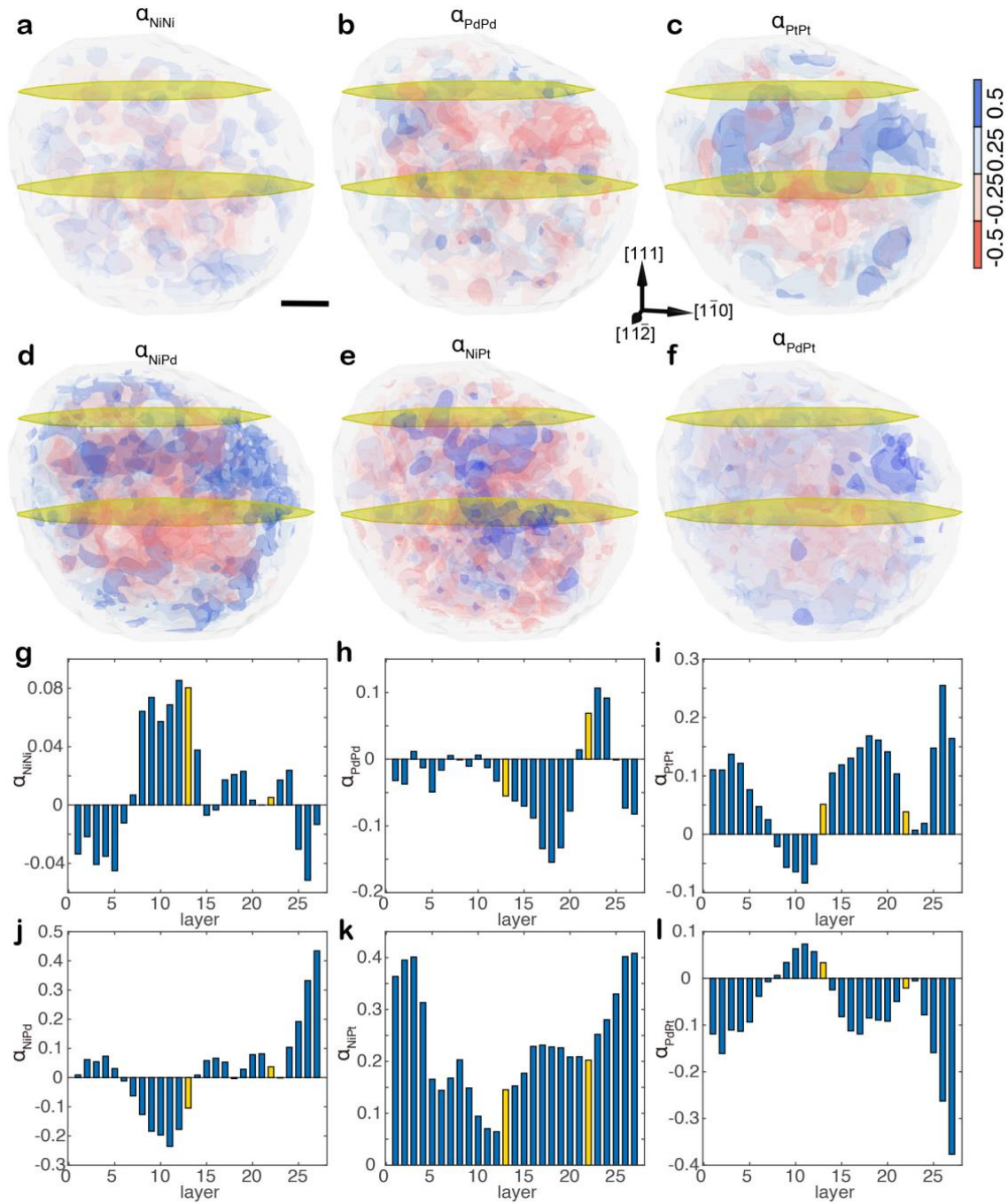

**Extended Data Fig. 6 | 3D distribution of the six CSRO parameters in the double-twinned MEA-3. a-f,** 3D distribution of $\alpha_{NiNi}$, $\alpha_{PdPd}$, $\alpha_{PtPt}$, $\alpha_{NiPd}$, $\alpha_{NiPt}$, and $\alpha_{PdPt}$, where the twins are labelled as yellow planes. **g-l**, Histogram of the average $\alpha_{NiNi}$, $\alpha_{PdPd}$, $\alpha_{PtPt}$, $\alpha_{NiPd}$, $\alpha_{NiPt}$, and $\alpha_{PdPt}$ values for each atomic layer along the [111] direction, where the yellow bars indicate the twin positions. Scale bar, 1 nm.

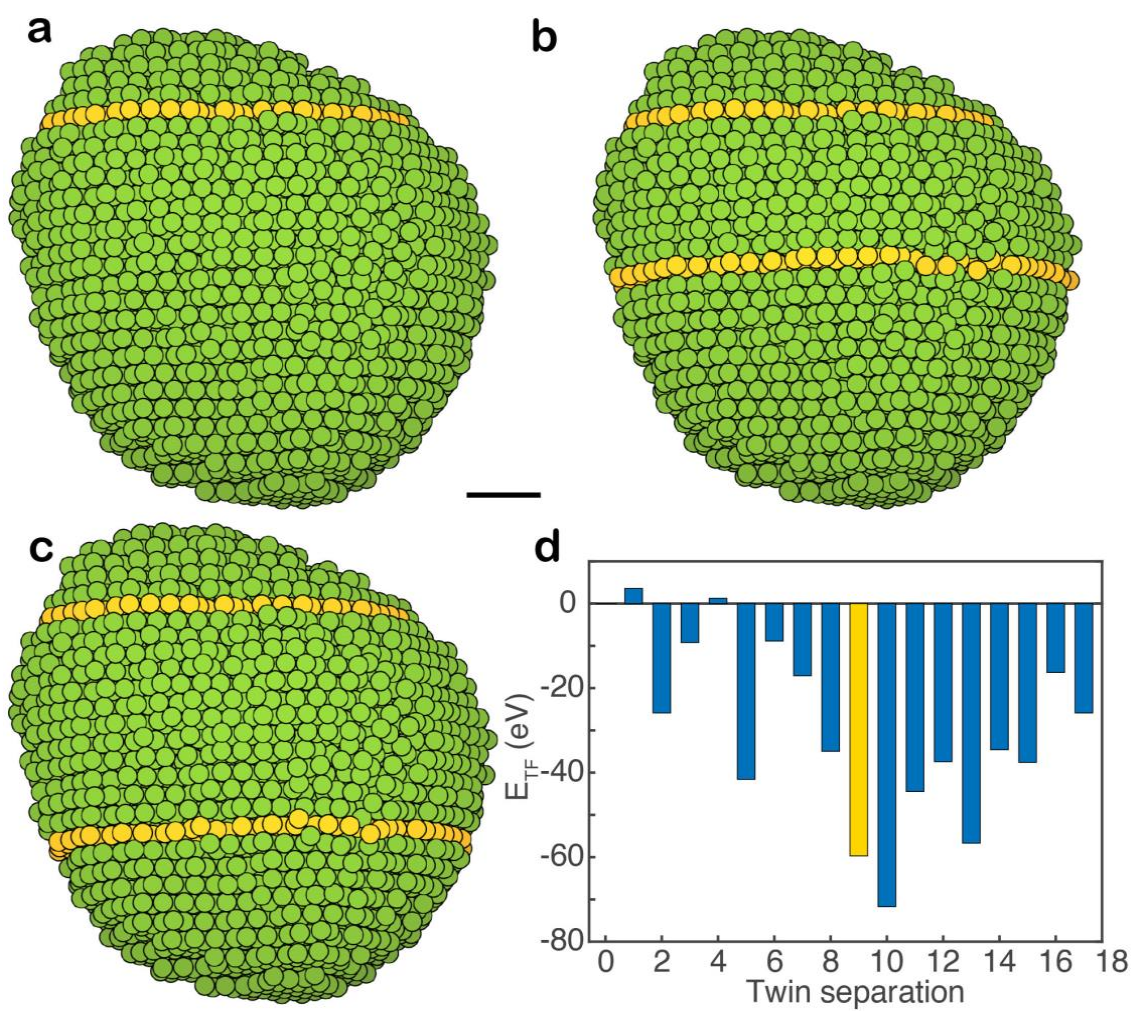

**Extended Data Fig. 7 | Twin formation energy ($E_{TF}$) calculated from experimental 3D atomic coordinates of the double-twinned MEA-3**. **a-c**, Calculation of $E_{TF}$ of the double-twinned MEA by fixing one twin (top yellow circles) and moving the other twin from the $0^{th}$ to the $14^{th}$ atomic layer along the [111] direction, where three representative atomic configurations show a twin separation by 0 (i.e., a single twin) (**a**), 9 (**b**), and 13 atomic layers (**c**). **d**, Histogram of $E_{TF}$ as a function of the twin separation. The experimentally determined twin separation is 9 atomic layers (yellow bar), which is next to the minimum $E_{TF}$ with a twin separation of 10 layers. Scale bar, 1 nm.

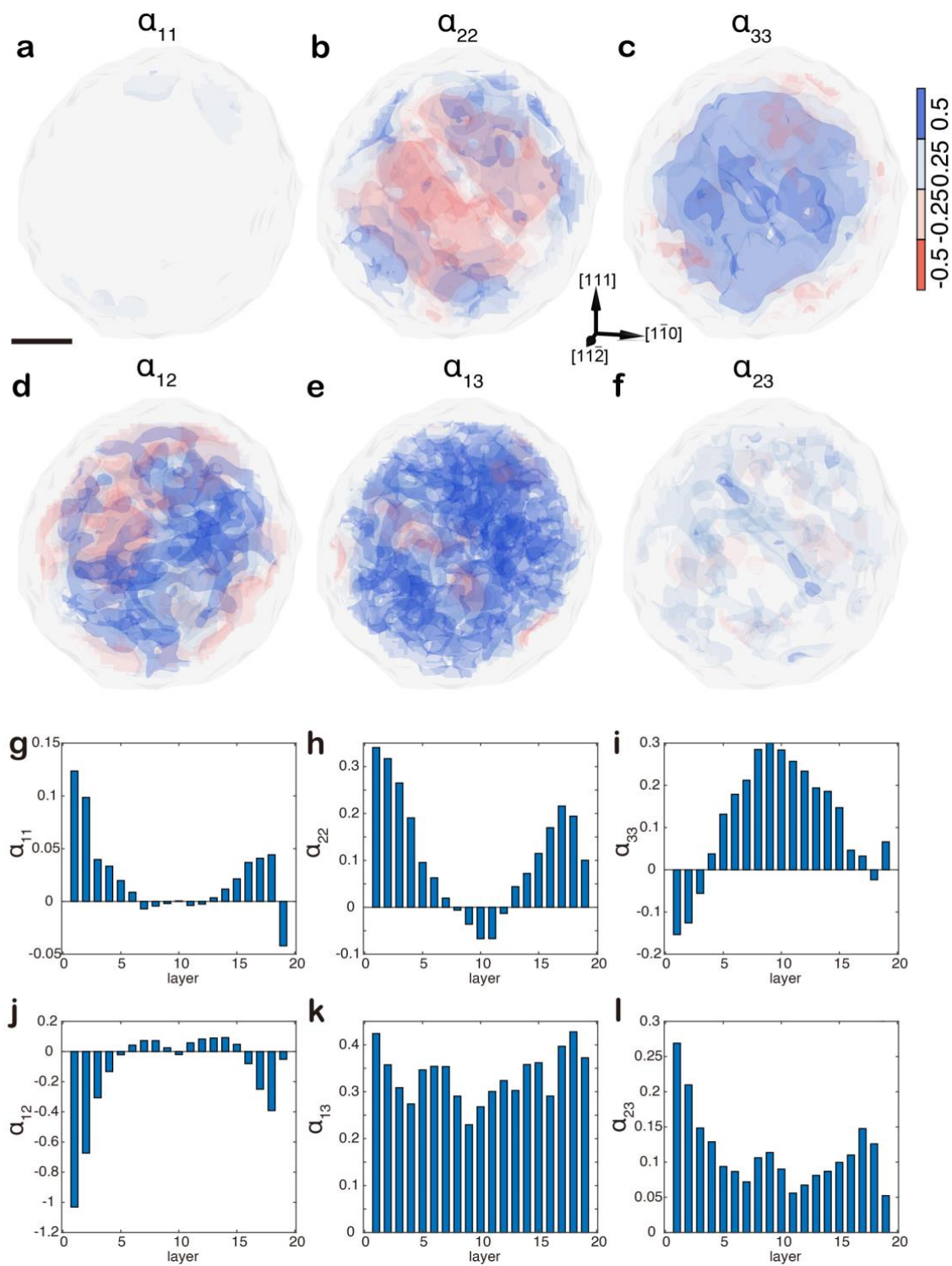

**Extended Data Fig. 8 | 3D distribution of the six CSRO parameters in the twin-free HEA-1**. **a-f**, 3D distribution of $\alpha_{11}$, $\alpha_{22}$, $\alpha_{33}$, $\alpha_{12}$, $\alpha_{13}$, and $\alpha_{23}$, which are more heterogeneous than those of the twin-free MEAs (Fig. 3a, b, Extended Data Fig. 5a-d). **g-l**, Histograms of the average $\alpha_{11}$, $\alpha_{22}$, $\alpha_{33}$, $\alpha_{12}$, $\alpha_{13}$, and $\alpha_{23}$ values for the atomic layer along the [111] direction. Scale bar, 1 nm.

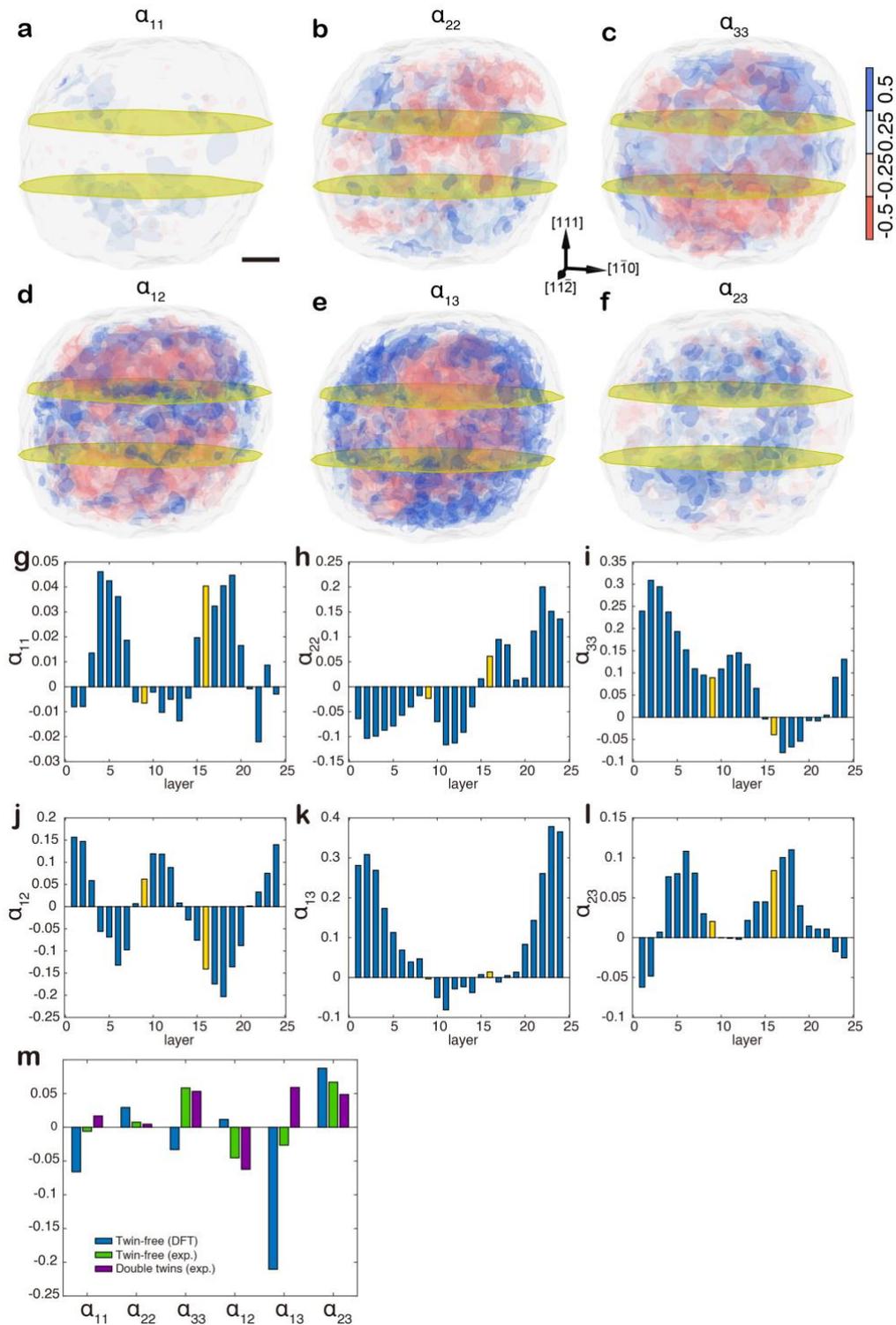

**Extended Data Fig. 9 | 3D distribution of the six CSRO parameters in the double-twinned HEA-2**. **a-f**, 3D distribution of $\alpha_{11}$, $\alpha_{22}$, $\alpha_{33}$, $\alpha_{12}$, $\alpha_{13}$, and $\alpha_{23}$, exhibiting greater local chemical fluctuations than the double-twinned MEA (Fig. 3e, f, Extended Data Fig. 5i-l and 6a-f)). **g-l**, Histograms of the average $\alpha_{11}$, $\alpha_{22}$, $\alpha_{33}$, $\alpha_{12}$, $\alpha_{13}$, and $\alpha_{23}$ values of the atomic layer along the [111] direction. **m**, Histogram of the six average CSRO parameters of a DFT-calculated bulk HEA, the twin-free HEA-1, and the double-twinned HEA-2. Scale bar, 1 nm.

**Extended Table 1 | AET data collection, processing, reconstruction, refinement and statistics of M/HEAs.**

|  | MEA-1[a] | MEA-2 | MEA3 | MEA-4 | MEA5 | MEA-6 | HEA-1 | HEA-2 | HEA-3 | HEA-4 |
|---|---|---|---|---|---|---|---|---|---|---|
| **Data collection and processing** | | | | | | | | | | |
| Voltage (kV) | 200 | 300 | 300 | 200 | 300 | 200 | 200 | 200 | 200 | 200 |
| Convergence semi-angle (mrad) | 25 | 30 | 30 | 25 | 30 | 25 | 25 | 25 | 25 | 25 |
| Probe size (Å) | 0.8 | 0.8 | 0.8 | 0.8 | 0.8 | 0.8 | 0.8 | 0.8 | 0.8 | 0.8 |
| Detector inner angle (mrad) | 38 | 38 | 38 | 38 | 38 | 38 | 38 | 38 | 38 | 38 |
| Detector outer angle (mrad) | 190 | 190 | 190 | 190 | 190 | 190 | 190 | 190 | 190 | 190 |
| Depth of focus (nm) | 14 | 14 | 14 | 14 | 14 | 14 | 14 | 14 | 14 | 14 |
| Pixel size (Å) | 0.347 | 0.434 | 0.434 | 0.347 | 0.306 | 0.347 | 0.347 | 0.347 | 0.347 | 0.347 |
| # of images | 57 | 53 | 54 | 56 | 59 | 55 | 56 | 61 | 54 | 57 |
| # of frames/angle | 3 | 3 | 3 | 3 | 3 | 4 | 4 | 3 | 3 | 3 |
| Tilt range (°) | -74.3° | -72.6° | -74.3° | -72.0° | -72.6° | -72.6° | -74.3° | -73.6° | -74.3° | -72.6° |
|  | 66.4° | 69.4° | 66.4° | 69.4° | 66.4° | 69.4° | 63.4° | 66.4° | 66.4° | 66.4° |
| Total electron dose ($10^5$ e/Å$^2$) | 7.9 | 7.4 | 7.5 | 7.8 | 8.2 | 7.7 | 7.8 | 8.5 | 7.5 | 7.9 |
| **Reconstruction** | | | | | | | | | | |
| Algorithm | RESIRE | RESIRE | RESIRE | RESIRE | RESIRE | RESIRE | RESIRE | RESIRE | RESIRE | RESIRE |
| Oversampling ratio | 4 | 4 | 4 | 4 | 4 | 4 | 4 | 4 | 4 | 4 |
| Number of iterations | 200 | 200 | 200 | 200 | 200 | 200 | 200 | 200 | 200 | 200 |
| **Refinement** | | | | | | | | | | |
| $R_1$ (%)[b] | 10.10 | 10.29 | 13.68 | 12.00 | 11.43 | 10.03 | 7.90 | 8.46 | 9.95 | 8.21 |
| R (%)[c] | 8.38 | 6.89 | 10.17 | 10.23 | 9.05 | 7.05 | 5.60 | 5.89 | 6.48 | 5.84 |
| B' factors (Å$^2$) | | | | | | | | | | |
| Ni / Type 1 | 42.30 | 38.82 | 33.11 | 29.04 | 54.51 | 42.34 | 41.63 | 45.77 | 33.95 | 35.5 |
| Pd / Type 2 | 37.95 | 28.47 | 25.68 | 29.74 | 37.46 | 37.87 | 37.62 | 46.41 | 32.81 | 34.5 |
| Pt / Type 3 | 29.95 | 22.88 | 24.55 | 25.29 | 30.65 | 26.97 | 36.33 | 40.92 | 29.86 | 34.6 |
| **Statistics** | | | | | | | | | | |
| # of atoms | | | | | | | | | | |
| Total | 2436 | 8349 | 8141 | 1794 | 5099 | 5323 | 5672 | 10053 | 3696 | 6037 |
| Ni / Type 1 | 989 | 2522 | 2399 | 545 | 1741 | 1514 | 1124 | 2638 | 872 | 1446 |
| Pd / Type 2 | 597 | 3247 | 3476 | 430 | 2010 | 1799 | 2356 | 3830 | 1735 | 2045 |
| Pt / Type 3 | 850 | 2580 | 2266 | 819 | 1348 | 2010 | 2192 | 3585 | 1089 | 2546 |

[a] MEA-1, MEA-2, HEA-1 and HEA-2 are the four nanocrystals shown in main text figures.
[b] The $R_1$-factor is defined as equation 5 in ref. 63.
[c] The R-factor is defined in equation 4 in ref. 58.